\documentclass[twocolumn,prb,showpacs,preprintnumbers,amsmath,amssymb]{revtex4-1}
\usepackage{graphicx}% Include figure files
\usepackage{dcolumn}% Align table columns on decimal point
\usepackage{bm}% bold math
\usepackage{verbatim}

\begin{document}

%\preprint{APS/123-QED}

\title{The diamagnetism above the superconducting transition in underdoped La$_{1.9}$Sr$_{0.1}$CuO$_4$ revisited: Chemical disorder or phase incoherent superconductivity?}

\author{Jes\'us Mosqueira}
\author{Javier D. Dancausa}
\author{F\'elix Vidal}
\affiliation{LBTS, Departamento de F\'isica da Materia Condensada, Universidade de Santiago de Compostela, E-15782 Spain}

\date{\today}

\begin{abstract}
The interplay between superconducting fluctuations and inhomogeneities presents a renewed interest due to recent works of different groups which apparently support an intrinsically anomalous (beyond the conventional Gaussian-Ginzburg-Landau scenario) diamagnetism above $T_c$ in underdoped cuprates. This conclusion, mainly based in the observation of new anomalies in the isothermal magnetization curves at low field amplitudes, is in contradiction with our earlier results in the underdoped La$_{1.9}$Sr$_{0.1}$CuO$_4$ [Phys.~Rev.~Lett.~\textbf{84}, 3157 (2000)]. These seemingly intrinsic anomalies are being presented in various influential works as a ``\textit{thermodynamic evidence}" for phase incoherent superconductivity in the pseudogap regime, this last being at present a central and debated issue of the cuprate superconductors' physics. To further probe the diamagnetism above $T_c$ in underdoped cuprates, here we have extended our magnetization measurements in  La$_{1.9}$Sr$_{0.1}$CuO$_4$ to two samples with the same nominal composition but, due to different growth procedures, with different chemical disorder, in one of the samples this disorder being close to the intrinsic-like one, associated with the unavoidable random distribution of the Sr ions (which will be then present even in an ideal La$_{1.9}$Sr$_{0.1}$CuO$_4$ crystal). For this sample, the corresponding $T_c$-inhomogeneities may be approximated as symmetric around the average $T_c$. In contrast, the most disordered sample presents a pronounced asymmetric $T_c$-distribution. The comparison between the magnetization measured in both samples provide a crucial check of the chemical disorder origin of the observed diamagnetism anomalies, which are similar to those claimed as due to phase fluctuations by other authors. This conclusion applies also to the sample affected only by the intrinsic-like chemical disorder, providing then a further check that, for all applied magnetic field amplitudes, the intrinsic diamagnetism above the superconducting transition of underdoped cuprates is not affected by the opening of a pseudogap in the normal state. It is also shown here that once these disorder effects are overcome, the remaining precursor diamagnetism may be accounted at a quantitative level in terms of the Gaussian-Ginzburg-Landau approach under a total energy cutoff.
\end{abstract}

\maketitle

\section{Introduction}

In high-$T_c$ cuprate superconductors (HTSC), the dilemma between superconducting fluctuations above the superconducting critical temperature, $T_c$,  and inhomogeneities was already posed by Bednorz and M\"uller in their seminal work:\cite{uno} after having indicated that the rounding of the electrical resistivity around the average $T_c$  in their lanthanum-barium-copper oxide could be due to inhomogeneities, they added  ``\textit{the onset} (of the resistivity drop) \textit{can also be due to fluctuations in the superconducting wave function}''. It was soon established that the observed rounding effects were much stronger than the ones expected from the presence of fluctuating superconducting pairs and that they could be explained by the presence of extrinsic  $T_c$-inhomogeneities with long characteristic lengths [larger than the in-plane superconducting coherence length amplitude, $\xi_{ab}(0)$], associated with chemical inhomogeneities.\cite{dos} In fact, their layered nature and the complexity of their chemistry enhance the relevance of the extrinsic inhomogeneity effects in real HTSC. Since these earlier results, the entanglement between superconducting fluctuations and inhomogeneities with different characteristic lengths and spatial distributions has played a central role when analyzing the measurements of any observable around the superconducting transition in cuprates.\cite{dos}
     
In the case of the rounding above $T_c$ of the normal state magnetization, the  so-called \textit{precursor diamagnetism}, the dilemma in cuprate superconductors between superconducting fluctuations above $T_c$ and inhomogeneities was open years ago, mainly in the underdoped compounds having a pseudogap in the normal state. Earlier measurements in the underdoped La$_{1.9}$Sr$_{0.1}$CuO$_4$ (LSCO-0.1) suggested that its precursor diamagnetism still is, from a phenomenological point of view, conventional.\cite{tres} In fact, not too close to $T_c$ these magnetization rounding effects were explained at a quantitative level in terms of the Gaussian-Ginzburg-Landau (GGL) approach for homogeneous layered superconductors.\cite{tres}  Therefore, these results suggested that, as it was also the case of the BCS metallic low-$T_c$ superconductors,\cite{cuatro,cinco,seis} or of the optimal doped cuprates,\cite{siete,ocho} the precursor diamagnetism in underdoped cuprates is due to fluctuating superconducting pairs, and is not affected by the opening of a pseudogap in the normal state. The anomalies observed since then by various groups in different HTSC, including the La$_{1.9}$Sr$_{0.1}$CuO$_4$  compound,\cite{nueve,diez} fueled different theoretical proposals of unconventional (non-GGL) precursor diamagnetism, in some cases seemingly confirming the popular scenario of phase incoherent superconductivity up to the pseudogap temperature.\cite{nueve,diez,once,doce}  However, these anomalies were later easily explained in terms of $T_c$ inhomogeneities with long characteristic lengths associated to chemical disorder, which do not directly affect the  superconducting transition own nature.\cite{trece1,trece2,trece3}  In fact, as the diamagnetism in the superconducting state is orders of magnitude stronger than the one associated with fluctuating superconducting pairs, even the $T_c$-inhomogeneities inherent to the unavoidable random distribution of dopants could have a dramatic influence on the measurements. The relevance of these intrinsic-like $T_c$-inhomogeneities on the measurements of the precursor diamagnetism was probed in recent experiments in the LSCO system.\cite{catorce}   
        
The debate about the nature of the precursor diamagnetism in underdoped cuprates has been recently reopened by new measurements and analyses, some of which were again performed in LSCO compounds, which seemingly confirm the intrinsically anomalous behavior of their magnetization above $T_c$.\cite{quince,dieciseis,diecisiete} These anomalies include a huge and field dependent diamagnetism close to the average $T_c$,  and an increase with the temperature of the so-called upturn magnetic field, $H_{up}$, at which the differential magnetic susceptibility changes from negative to positive on increasing the applied magnetic field, a behavior that cannot be explained by a Gaussian $T_c$ distribution.\cite{quince}  These different magnetization anomalies are being considered in various influential works as an evidence for phase incoherent superconductivity well above $T_c$,\cite{dieciocho,diecinueve,veinte} in particular because ``\textit{there is hardly another known mechanism that can deliver diamagnetism of this magnitude}".\cite{diecinueve} The relevance of the precursor diamagnetism is enhanced by the fact that another central experimental support for phase incoherent superconductivity above $T_c$ in cuprates, the relatively large amplitude of the Nernst effect well above $T_c$, has been recently questioned by various groups, which attributed such behavior to normal state effects.\cite{veintiuno,veintidos} So, at present one of the main experimental arguments invoked to still support the popular phase-disordering scenario above $T_c$ for the cuprate superconductors is the seemingly non-GGL precursor diamagnetism.\cite{dieciocho,diecinueve,veinte, veintitres} In fact, the coincidence of these seemingly intrinsic anomalies with those observed in measurements of the Nernst effect has been claimed as ``\textit{a fairly unequivocal diagnosis of vortex motion}" above $T_c$.\cite{dieciocho}
   
The results summarized above stress the interest of further study the influence of different inhomogeneities on the precursor diamagnetism in underdoped cuprate superconductors, in particular complementing our previous works on that issue.\cite{trece1,trece2,trece3,catorce} This is the central aim of our present paper, where we will first present magnetization measurements in two underdoped granular samples of the same nominal composition, La$_{1.9}$Sr$_{0.1}$CuO$_4$ (LSCO-0.1), but with different structural and chemical disorder due to differences in the synthesis. These differences lead, in particular, to different x-ray diffraction linewidths and superconducting transition widths. 
These preliminary characterization measurements show that in one of the samples the chemical disorder is close to the intrinsic-like one, associated with the unavoidable random distribution of the Sr ions (which will be then present even in an ideal La$_{1.9}$Sr$_{0.1}$CuO$_4$ crystal). The chemical disorder in this sample leads to a (intrinsic-like) $T_c$-distribution which may be approximated as symmetric, whereas the most disordered sample present a pronounced  asymmetric $T_c$-distribution. 
As we will see here, a direct comparison between the as-measured magnetization in both samples already demonstrates that most of the observed anomalies are in fact extrinsic.  When analyzed at a quantitative level, these data show that the anomalies at low field amplitudes in the isothermal magnetization curves may be explained by the presence of structural and $T_c$-inhomogeneities with long characteristic lengths, much larger than $\xi_{ab}(0)$. This conclusion includes both the upturn magnetic field, $H_{up}$,\cite{nueve,diez,quince,dieciseis} and the apparently anomalous power-law dependence below $H_{up}$, a behavior pompously called by some authors ``\textit{fragile London rigidity}".\cite{diez,dieciseis} 
     
Our results will also provide a direct explanation in terms of $T_c$-inhomogeneities of a quite subtle anomaly, already commented above, affecting these isotherms: The increase of $H_{up}$ with the temperature, a dependence recently observed by Lascialfari and co-workers\cite{quince} at a quantitative level and that cannot be explained in terms of a Gaussian $T_c$ distribution.  It will be shown here that this behavior may be still attributed to chemical disorder if it leads to an asymmetric $T_c$-distribution, extending well above the average transition temperature, $\overline T_c$. In fact, this is mainly the case in the most inhomogeneous samples, where such a temperature dependence of  $H_{up}$ is observed to be more important: even a symmetric doping-level distribution could lead to an asymmetric $T_c$-distribution, due to the characteristic bell-shaped $T_c$ dependence on the doping level, $x$. The only requisite is that such doping-level distribution is wide enough as to cover a region where $T_c$ presents a strongly non-linear $x$-dependence. This will be the case of the underdoped LSCO if this region extends up to the optimum doping. 

Then, it is shown that the inhomogeneity and the fluctuation effects may be disentangled by just applying a magnetic field large enough as to shift $\overline T_c(H)$ to low temperatures, by an amount of the order or larger than the superconducting transition width. In this way, we were able to quench inhomogeneity effects in the less inhomogeneous sample. The remaining diamagnetism is explained, even at a quantitative level, in terms of the GGL approach for homogeneous layered superconductors.  Complementarily, this agreement with the GGL approach also confirms our previous conclusions about the absence in the bulk of inhomogeneities with short characteristic lengths, as those observed by using surface probes.\cite{veinticuatro} Indirectly, they also support recent proposals that the large-amplitude Nernst signal observed in the normal state in LSCO is not associated with superconducting fluctuations.\cite{veintiuno,veintidos}

The paper is organized as follows: The fabrication and characterization of the samples is described in Sec.~II. The experimental results on the low-field anomalies observed around $\overline T_c$ are presented in Sec.~III.A, while in Sec.~III.B these anomalies are explained in terms of an inhomogeneity model which takes into account the $T_c$ distributions present in the samples. Sec.~IV is dedicated to the data analyses in terms of the GGL approach for layered superconductors, paying special attention to the reduced-temperature dependence of the precursor diamagnetism, which further confirm the presence of $T_c$-inhomogeneities. The conclusions are presented in Sec.~V.

\section{Fabrication and general characterization of the samples}

%
%
% Fig1
%
%
\begin{figure}[t]
\includegraphics[scale=1.3]{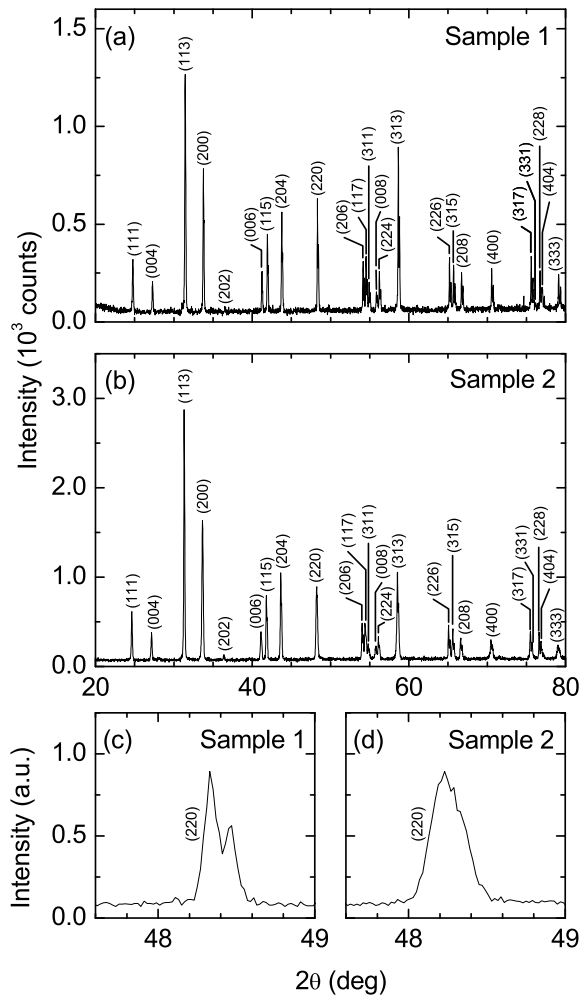}
\caption{(a) and (b) X-ray diffraction patterns for sample 1 and 2, respectively.  The positions of the diffraction peaks of both samples coincide, indicating that they have the same average chemical composition. However, those corresponding to sample 2 (with less grind-react processes) present larger linewidths, providing a first indication of a more inhomogeneous Sr distribution. This is illustrated in (c) and (d), which show details of the Cu-$K_{\alpha 1}$ and the Cu-$K_{\alpha 2}$ components of the (220) line for sample 1 and 2, respectively. Whereas in sample 2 the two components overlap, in the most homogeneous sample they are resolved.}
\label{rx}
\end{figure}

We have used two granular La$_{1.9}$Sr$_{0.1}$CuO$_4$ (LSCO-0.1) samples, that were chosen instead of single crystals because they allow to change easily the structural and $T_c$-inhomogeneities, but conserving the same average chemical composition.  This is a crucial experimental aspect of this work.  In addition, the random orientation of the grains and the influence of the grains finite size may be easily taken into account when analyzing their magnetic behavior. Following a standard procedure described in detail elsewhere (see, e.g., Ref.~\onlinecite{catorce} and references therein), both samples were prepared by reacting in air at 930$^\circ$C for 20 h stoichiometric proportions of thoroughly mixed powders of La$_2$O$_3$, SrCO$_3$, and CuO (99.99\% purity). Sample 1 was reacted up to ten times with intermediate grindings in an agate mortar. Such a number of grind-react (G-R) processes was shown to be enough to attain a superconducting transition-width close to the \textit{intrinsic} one (associated to the random distribution of the Sr dopants).\cite{catorce} However, to study the influence of the extrinsic structural and chemical inhomogeneities on the precursor diamagnetism, sample 2 was prepared in a slightly different way: on the one side, it has been subjected to only four G-R process, so one may expect that this sample will have more chemical inhomogeneities and then more $T_c$-inhomogeneities. 
On the other, in order to obtain a smaller grain size, the intermediate grindings were performed with a commercial grinder (Retsch, model PM 100). As we will see below, transition width and grain size are both factors strongly affecting the magnetic behavior of these materials around the average~$T_c$.

A first check of the samples' chemical composition and homogeneity was provided by x-ray diffraction (XRD) measurements performed with a Philips diffractometer equipped with a Cu anode and a Cu-K$_\alpha$ graphite monocromator.  As it may be seen in Fig.~\ref{rx}, the XRD patterns of both samples exclude the presence of impurity phases and are almost indistinguishable, the only difference being the larger linewidths in the case of sample 2 (fabricated with less G-R processes). This behavior related to the number of G-R processes was already reported in Ref.~\onlinecite{catorce} and may be attributed to a more inhomogeneous Sr distribution in sample 2, which directly leads, as we will see below, to a wider $T_c$ distribution. Another relevant difference between both samples concerns their average grain diameter, 5 $\mu$m for sample 1 and  1 $\mu$m for sample 2 as determined by scanning electron microscopy.

%
%
% Fig2
%
%
\begin{figure}[!b]
\includegraphics[scale=0.3]{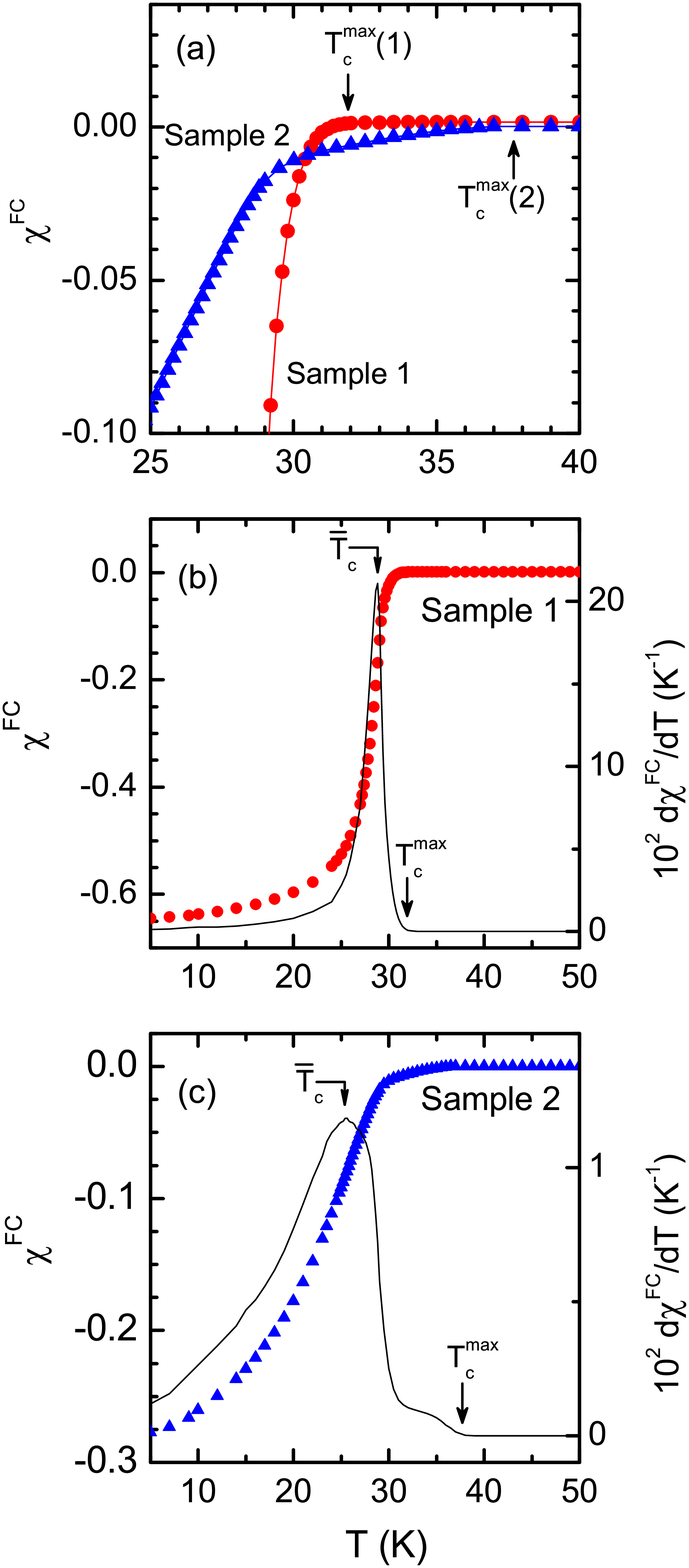}
\caption{(Color online) Temperature dependence of the field-cooled magnetic susceptibility, $\chi^{FC}$, measured under very low magnetic field amplitudes (0.5 mT). The different behavior around $\overline T_c$ in the two samples studied here may be observed in (a): whereas sample~1 presents a relatively well defined diamagnetic transition, sample~2 is diamagnetic up to a temperature, $T_c^{\rm max}$,  close to the highest critical temperature of the La$_{2-x}$Sr$_x$CuO$_4$ system ($\sim38$ K). An overview of these $\chi^{FC}(T)$ curves, together with their temperature derivatives (solid lines), are shown in (b) and (c). One may see in (c) that the transition width, $\Delta T_c$, for sample 2 does not take into account its markedly asymmetric $T_c$-distribution. Other details are commented in the main text.}
\label{tc}
\end{figure}
    
The average transition temperatures ($\overline T_c$) and transition widths ($\Delta T_c$) were estimated from the temperature dependence of the field-cooled (FC) magnetic susceptibility ($\chi^{FC}$) under low applied magnetic fields (see Table~I). $\overline T_c$ is defined as the temperature at which the temperature derivative of the FC curve has its maximum. $\Delta T_c$ is defined as twice the high-temperature half-width at half-maximum of the $d\chi^{FC} /dT$ versus $T$ curve. In this way we elude the extrinsic rounding associated with the competition between the grains size and the magnetic penetration length, which is appreciable mainly below $\overline T_c$ (see the note in Ref.~37 of Ref.~\onlinecite{trece1}). Also, as in this region $|\chi^{FC}|\ll1$, the demagnetization effects may be neglected. In the case of samples with a markedly asymmetric $T_c$-distribution, extending well above $\overline T_c$ (in a temperature region where $d\chi^{FC}/dT$ around $\overline T_c$ can no longer be approximated as Gaussian), as a complementary characterization of the $T_c$ distribution we introduce $T_c^{\rm max}$, roughly defined as the temperature above which $d\chi^{FC}/dT$ is below the experimental uncertainty.

%
%
% Fig3
%
%
\begin{figure}[b]
\includegraphics[scale=.3]{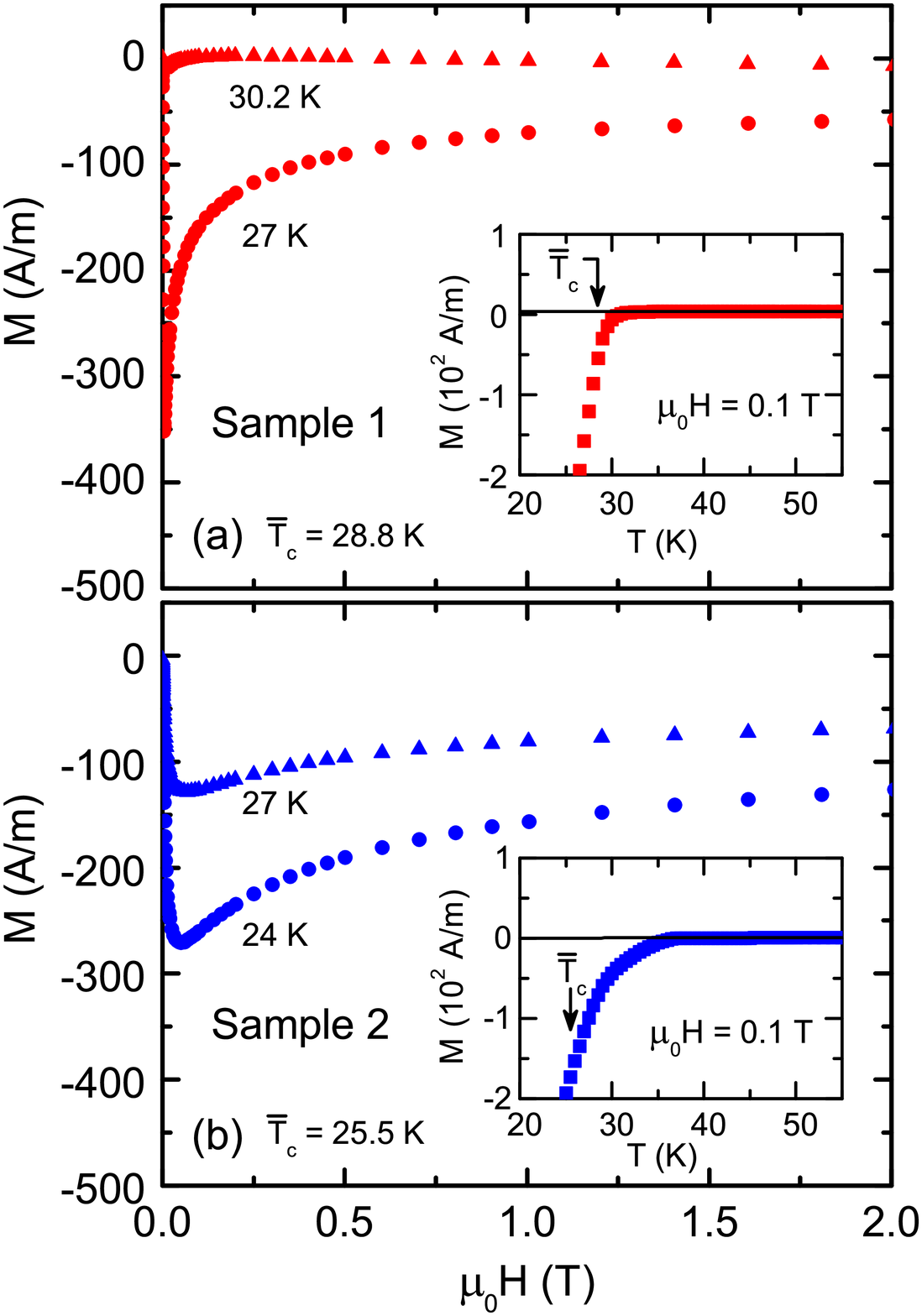}
\caption{(Color online) Some examples of the as-measured magnetization isotherms obtained in the two samples studied here under low-magnetic field amplitudes, $H/H_{c2}(0)\ll 1$. One may see that the behavior of these isotherms differs dramatically from sample to sample, mainly below $\overline T_c$.  The pronounced minimum at very low fields corresponds to the lower critical field ($H_{c1}$). In sample~2 one may observe a shift of this minimum to higher fields and also a rounding. Both behaviors clearly suggest the presence of $T_c$-inhomogeneities, which will affect more deeply sample~2. This scenario is supported by the as-measured magnetization versus temperature curves shown in the insets: the $M(T)_H$ curve measured in sample~2  is much more rounded above its $\overline T_c$ than the equivalent curve measured in sample~1. The solid lines here are the background magnetization, obtained as explained in the main text.}
\label{isotherms}
\end{figure}

An example of the $\chi^{FC}(T)$ curves for each sample is presented in Fig.~\ref{tc}, together with the corresponding temperature derivatives. These measurements were performed with a superconducting quantum interference magnetometer (Quantum Design, model MPMS-XL) and under low fields (around 0.5 mT). Important differences between these curves for both samples are readily observable in spite of the above mentioned similarities between their XRD patterns: On the one side, as a consequence of its larger average grain size, sample 1 presents a much larger Meissner signal. On the other, the improved Sr distribution in sample 1 makes its diamagnetic transition much sharper than in sample 2: for sample 1, $\Delta T_c/\overline T_c\approx0.04$ (which is close to the intrinsic value associated to the unavoidable random distribution of the Sr dopants, see Ref.~\onlinecite{catorce}), while $\Delta T_c/\overline T_c\approx0.25$ for sample~2. In fact, as may be already observed in Fig.~\ref{tc}(a), this sample is slightly diamagnetic up to a temperature near the maximum transition temperature for the La$_{2-x}$Sr$_x$CuO$_4$ system ($\sim38$~K), see Table I. Such a behavior is illustrated at a more quantitative level in Figs.~\ref{tc}(b) and (c) through the  $d\chi^{FC}/dT$ curves.  We will see in the next Section that such an asymmetric $T_c$-distribution plays a crucial role in the fine-detail behavior of the precursor diamagnetism measured in the most inhomogeneous samples. In particular, it leads to an increase of $H_{up}$ as the temperature of the corresponding isotherm increases. The values of these different parameters are summarized in Table I.

\section{Magnetization isotherms in the low field regime around the average transition temperature}

As already stressed in the Introduction, some of the magnetization anomalies claimed in different works as convincing experimental evidences of an unconventional superconducting transition in cuprates, were observed in the low-field regime of the magnetization isotherms above the transition temperature.\cite{nueve,quince,dieciseis,diecisiete,dieciocho,diecinueve} These seemingly intrinsic anomalies include the presence at low field amplitudes of the upturn magnetic field, $H_{up}$, its temperature dependence, and a field-dependent and huge diamagnetism below $H_{up}$, orders of magnitude larger than the diamagnetism associated with the presence of fluctuating superconducting pairs. So, in this Section we will first present detailed measurements of these isotherms in the two samples studied here. We will directly see in the as-measured curves two central aspects: even at a qualitative level, the behavior around $\overline T_c$ at low fields is strongly sample dependent. This already suggests that such a behavior is not intrinsic. Moreover, in the most inhomogeneous sample 2, the magnetization around $H_{up}$ is of the same order of magnitude above and below $\overline T_c$, in both cases orders of magnitude larger than one may expect from the presence of fluctuating superconducting pairs.\cite{trece1,trece2,trece3,catorce} As already shown when analyzing other magnetization anomalies in other experiments,\cite{trece1,trece2,catorce} the presence of inhomogeneities like the ones revealed in our samples by x-ray diffraction and low-field magnetometry, could lead to so huge amplitudes above $\overline T_c$.

To characterize at a quantitative level the magnetization anomalies we will introduce the so-called excess magnetization, 
\begin{equation}
\Delta M(T,H)=M(T,H)-M_B(T,H),  
\label{Back}                 
\end{equation}
where $M_B(T,H)$ is the \textit{background magnetization} (free from rounding effects such as the intrinsic fluctuation effects) which may be approximated by extrapolating through the transition the normal state magnetization measured well above $T_c(H)$.

\subsection{Experimental results and some qualitative considerations about the observed anomalies on the magnetization isotherms}

Some examples of the as-measured magnetization isotherms obtained in the low field regime around $\overline T_c$ are shown in Fig.~\ref{isotherms}. This regime corresponds to $H/H_{c2}(0)\ll 1$, where  $H_{c2}(0)$ is the upper critical field perpendicular to the crystallographic $ab$ planes linearly extrapolated to  $T= 0$~K. In this limit, the conventional GGL approach for the excess-magnetization associated with the presence of fluctuating superconducting pairs above the superconducting transition reduces to the so-called Schmidt regime, where the precursor diamagnetism is proportional to the applied magnetic field.\cite{tres,cinco}  These measurements were performed by using again a high resolution magnetometer based on the superconducting quantum interference (Quantum Design, model MPMS-XL), and following the procedures described in detail elsewhere (see Refs.~\onlinecite{trece1,trece3,catorce}, and references therein). Let us just stress here that special care was taken to avoid spurious contributions to the measured magnetization, in particular those that could be associated with the presence in the sample holder of minute quantities of oxygen, which has a paramagnetic transition around  45 K, a temperature region particularly sensitive for the precursor diamagnetism extraction in LSCO compounds. 

%
%
% Fig4
%
%
\begin{figure}[b]
\includegraphics[scale=.3]{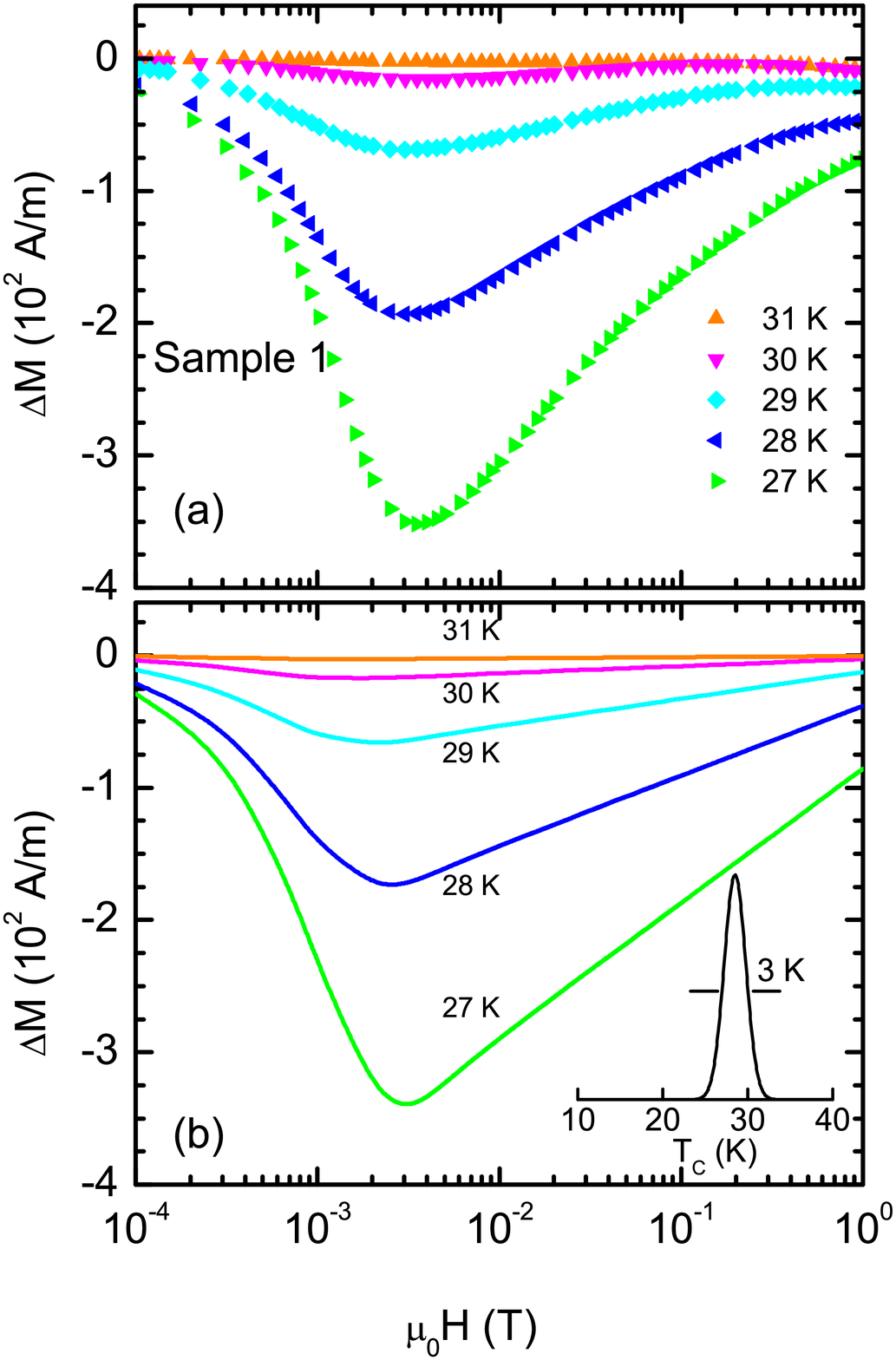}
\caption{(Color online) (a) Magnetic field dependence of the excess magnetization of sample~1 for several constant temperatures around $\overline T_c$. The background or normal-state contribution was removed from these curves by subtracting the $M(H)$ curve measured at 37~K, which is well above $\overline T_c$ and where the fluctuation diamagnetism is expected to be negligible in the scale of these figures. (b) The lines correspond to Eq.~(\ref{Integralangular}) numerically evaluated by using a Gaussian $T_c$ distribution with $\overline T_c=28.5$ K and $\Delta T_c=3$ K (represented in the inset in arbitrary units). Note that for this sample, with a symmetric $T_c$-distribution, the upturn magnetic field $H_{up}$ decreases as the temperature of the corresponding isotherm increases. }
\label{sample1}
\end{figure}

\begin{table}[!b]
\begin{ruledtabular}\begin{tabular}{cccccc}
Sample & $\overline T_c$ & $\Delta T_c$ & $T_c^{\rm max}$ & $\mu_0H_{c2}(0)$ & $\xi_{ab}(0)$ \\
& (K) & (K) & (K) & (T) & (\r{A}) \\
\hline
1 & 28.8 & 1.2 & 32 & 36 & 30 \\
2 & 25.5 & 6.8 & 38 & 25 & 36 \\
\end{tabular}\end{ruledtabular} 
\caption{Values of the main parameters arising in the phenomenological description of the diamagnetism above the superconducting transition measured in the two LSCO samples studied here (see main text for details). }
\end{table}

As already noted above, one may directly see in Fig.~\ref{isotherms} the dramatic differences between the magnetization isotherms measured in the two samples, mainly above $\overline T_c$.  This is so in spite that the isotherms in the two samples correspond to quite similar reduced-temperature  \textit{distances} to the transition, $|\varepsilon|\equiv |\ln(T/\overline T_c)|\approx|T-\overline T_c|/\overline T_c$ being around 0.06  for the four isotherms above and below the corresponding $\overline T_c$. As expected, below $\overline T_c$  both isotherms present a deep minimum associated with the lower critical field, $H_{c1}$, but this minimum is clearly smaller and rounded in sample~2, and also shifted to higher field amplitudes. This behavior clearly suggest the presence of $T_c$-inhomogeneities affecting more deeply sample~2. This scenario is supported by the as-measured magnetization versus temperature curves shown in the insets of Figs.~\ref{isotherms}(a) and (b): The $M(T)_H$ curve measured in sample~2  is much more rounded above its $\overline T_c$ than the equivalent curve measured in sample 1. The solid lines are the corresponding background magnetizations (for the details see next section).

\begin{figure}[t]
\includegraphics[scale=.3]{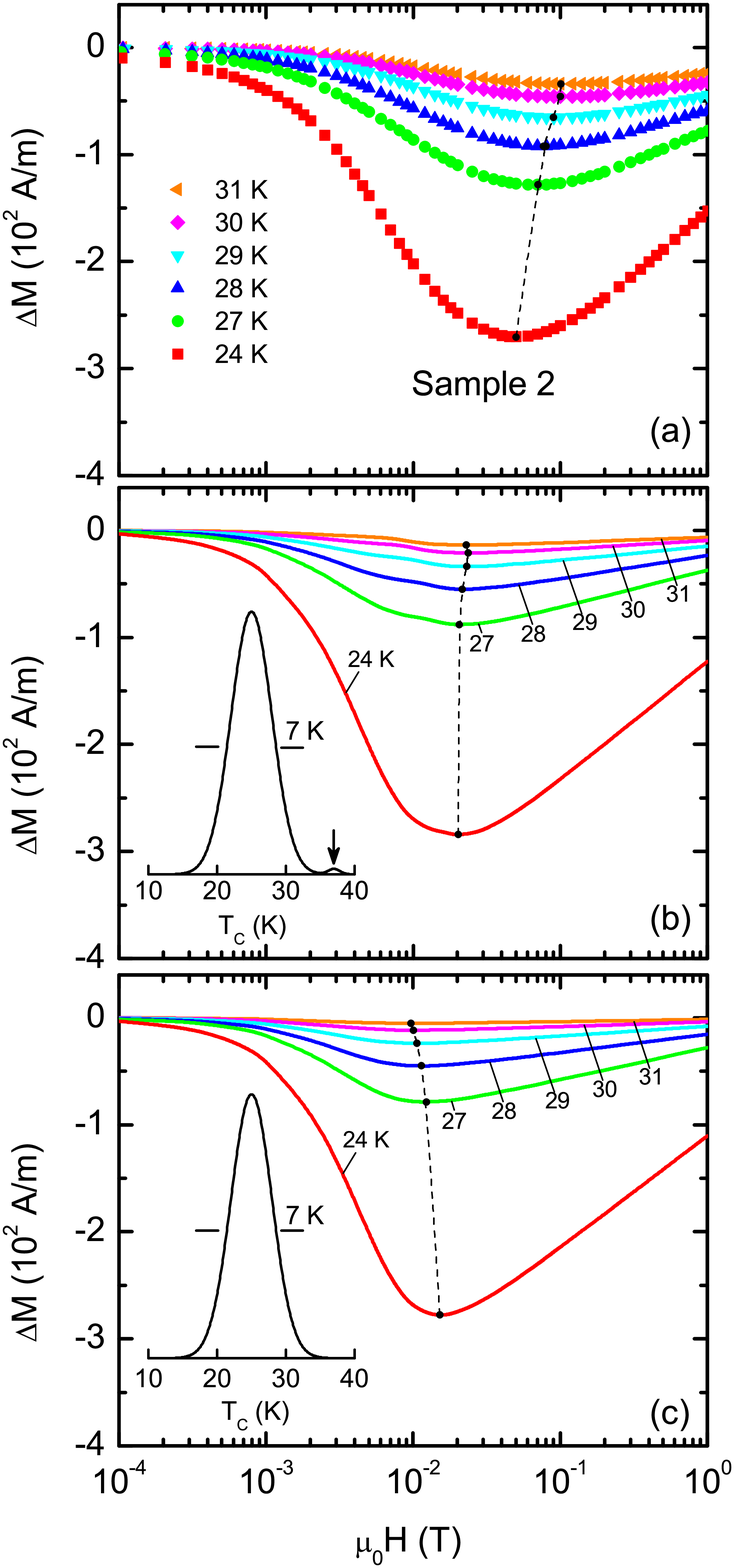}
\caption{(Color online) (a) Magnetic field dependence of the excess-magnetization of sample~2 for some constant temperatures around $\overline T_c$. These data were corrected for the normal-state contribution by subtracting the isotherm measured at 37~K (well above $\overline T_c$). The dashed line joins the data points of the different isotherms where $H_{\rm up}$ is located. In contrast with the results for sample 1 (Fig.~\ref{sample1}), in this sample $H_{up}$ incresases when the temperature of the corresponding isotherm increases. (b) and (c): The lines correspond to Eq.~(\ref{Integralangular}) evaluated with the $T_c$ distribution shown in the corresponding insets in arbitrary units. As may be clearly deduced by comparing the curves in (b) and (c), the $H_{\rm up}$ increase is a consequence of the slightly asymmetric $T_c$ distribution shown in the inset in (b), which has a higher-$T_c$ bump signaled by an arrow (see main text for details). }
\label{sample2}
\end{figure}

The differences in the as-measured magnetization observed in Fig.~\ref{isotherms} for the two samples may be better appreciated by comparing the results presented in Figs.~\ref{sample1} and \ref{sample2} for the excess magnetization isotherms, $\Delta M(H)_T$,  for temperatures just above $\overline T_c$ in the two samples studied here. In this case, the background contribution was estimated as the isotherm at 37~K, well above the corresponding $\overline T_c$. Both samples present a strongly nonlinear behavior: $|\Delta M(H)_T|$ increases with $H$ following a power law (see below) up to the \textit{upturn} magnetic field, $H_{up}$, and decreases monotonously for larger fields.  It is just the anomalous behavior at very low reduced field amplitudes first observed by Lascialfari and coworkers.\cite{nueve}  This anomaly, which cannot be explained on the grounds of  the Ginzburg-Landau  approaches for the diamagnetism induced by the presence of fluctuating superconducting pairs,\cite{cuatro} is  being claimed by different authors as an evidence of the relevance of phase fluctuations above $T_c$ in cuprate superconductors.\cite{nueve,quince,dieciseis,diecisiete,dieciocho,diecinueve} However, as already indicated in the Introduction, contrary to these proposals it was already shown in Refs.~\onlinecite{trece1,trece2,catorce} that the presence in the samples of $T_c$ inhomogeneities at long length scales (much larger than the coherence length amplitude) could lead to a similar $M(H)_T$ dependence around $\overline T_c$. In fact, a similar effect was observed also in disordered low-$T_c$ metallic alloys.\cite{trece1}

%
%
% Fig5
%
%

The results summarized in Figs.~\ref{sample1}(a) and \ref{sample2}(a) also allow to perform a crude but direct analysis of the observation that in some high-$T_c$ cuprates $H_{up}$ increases with temperature.\cite{nueve} As already commented in the Introduction, such a behavior is being presented as a strong argument against the explanation of the magnetization anomalies in terms of $T_c$-inhomogeneities:\cite{nueve}  in a naive inhomogeneity scenario,  $H_{up}$ will mimic the lower critical field of the higher-$T_c$ domains, which decreases on increasing the temperature. Nevertheless, the extrinsic nature of this effect is directly inferred by the fact that $H_{up}$ increases with temperature only in sample 2, and also its corresponding $H_{up}$ values differ by almost one order of magnitude from those of sample 1. Moreover, the fact that  sample 2 is just the one fabricated with a smaller number of grind-react processes and the one having a wider diamagnetic transition, strongly suggests already that such an $H_{up}$(T) dependence is associated with inhomogeneities, although in a way not so direct as for sample 1.  In fact, we will see in the next Section that such a behavior may be quantitatively explained in terms of an asymmetric $T_c$ distribution, extended well above the average $T_c$. In turn, this kind of distribution may be easily explained in terms of a symmetric doping-level distribution, due to the characteristic bell-shaped $T_c$ dependence on the doping level. The only requisite is that such doping-level distribution is wide enough as to be appreciable near the optimum doping. In this way domains near the maximum $T_c$ (where $T_c$ is almost independent of the hole content) may present a significant volume fraction even if the average hole content is well inside the underdoped or overdoped regions.

%
%
% Fig6
%
%
\begin{figure}[!b]
\includegraphics[scale=0.35]{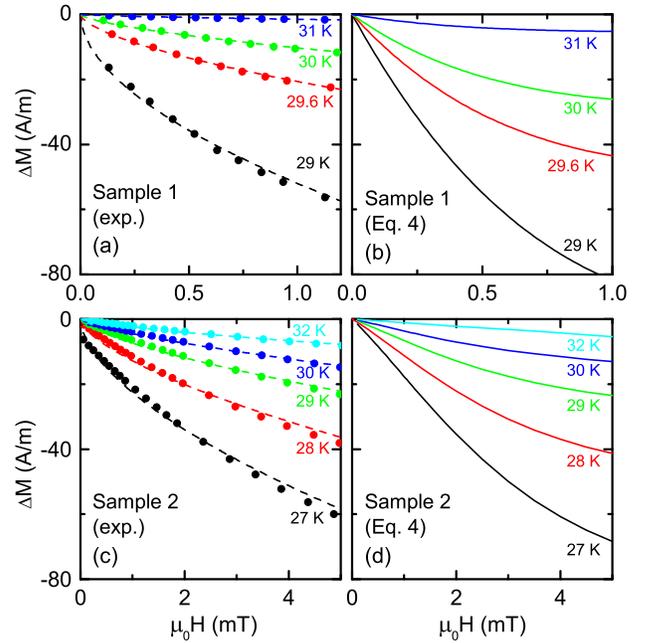}
\caption{(Color online) (a) and (c): Detail of the $\Delta M(H)$ dependence when $H<H_{up}$ for temperatures just above $\overline T_c$ in the two samples studied here. The dashed lines are fits of a power law [Eq.~(\ref{Power})]. As shown in (b) and (d), in spite of the simplifications introduced, the model developed in Section III.B for the effect of a $T_c$ distribution [Eq.~(\ref{Integralangular}), solid lines] roughly accounts for the experimental observations.}
\label{Lineal} 
\end{figure}

\subsection{Quantitative analysis of the isothermal magnetization curves above the superconducting transition in terms of the inhomogeneity scenario}

To calculate the magnetization above the superconducting transition in presence of $T_c$-inhomogeneities at long length scales, much larger than $\xi(0)$, we will use a simple approach similar to the one proposed in previous works.\cite{trece1,catorce} In this model the measured magnetization is approximated by the average
\begin{equation}
\left\langle \Delta M(T,H)\right\rangle=\int_0^{T_c^{\rm max}} dT_c\;\delta(T_c)\Delta M(T,H,T_c),
\label{Integral}
\end{equation}
where $\delta(T_c)$ represents the volume distribution of critical temperatures, and $\Delta M(T,H,T_c)$ the \textit{intrinsic} magnetization of an homogeneous superconducting domain having a critical temperature $T_c$. 
As we are dealing with polycrystal samples with the grains randomly oriented, we must angular average $\Delta M$. For that, the magnetization of an arbitrarily oriented crystallite in the direction of the applied magnetic field is approximated by
\begin{equation}
\Delta M_\theta(T,H,T_c)\approx \Delta M_\perp(T,H\cos\theta,T_c)\cos\theta,
\label{Mtheta}
\end{equation}
where $\theta$ is the angle between the applied magnetic field and the grain's $c$ axis, and $\Delta M_\perp$ is the component of the magnetization vector in the direction perpendicular to the CuO$_2$ ($ab$) layers. Note that we have neglected the contribution coming from the parallel component of the magnetization vector due to the large anisotropy of these compounds ($\gamma\sim$10-15, see, e.g., Ref.~\onlinecite{veinticinco}). With the above approximations, the effective magnetization of a polycrystal in presence of $T_c$ inhomogeneities is given by
\begin{eqnarray}
&&\!\!\!\!\!\!\langle \Delta M(T,H)\rangle\approx\nonumber\\
&&\!\!\!\!\!\!\int_0^{\pi/2}\!\!\!\!\!\! d\theta\sin\theta\int_0^{T_c^{\rm max}}\!\!\!\!\!\!\!\!\Delta M_\perp(T,H\cos\theta,T_c)\cos\theta\;\delta(T_c)dT_c.
\label{Integralangular}
\end{eqnarray}
In this equation, the intrinsic perpendicular magnetization was obtained through 
\begin{equation}
\Delta M_\perp(T,H,T_c)=m(t,h)H_{c2}(0),
\label{mred}
\end{equation}
where $t\equiv T/T_c$ is the normalized temperature, $h\equiv H/H_{c2}(0)$ the reduced magnetic field, and the reduced magnetization $m(t,h)$ may be approximated as follows (see, e.g., Ref.~\onlinecite{veintiseis}):

i) In the \textit{Meissner region}, i.e., for $h/(1-t)<\ln\kappa/2\kappa^2$, where $\kappa$ is the Ginzburg-Landau parameter
\begin{equation}
m=-\alpha\frac{h}{1-D}.
\label{mMeissner}
\end{equation}
Here $D$ is the demagnetizing factor and $\alpha$ accounts for a reduction from the ideal diamagnetic response caused by a competition between the grains size and the \textit{in-plane} magnetic penetration length, $\lambda_{ab}$, which for this material is about 0.3 $\mu$m when $T\to0$ K.\cite{veinticinco} In the case of spherical grains $D=1/3$ and, according to Ref.~\onlinecite{veintisiete}
\begin{equation}
\alpha=1-\frac{3\lambda_{ab}(T)}{r}\coth\left[\frac{r}{\lambda_{ab}(T)}\right]+\frac{3\lambda_{ab}^2(T)}{r^2},
\label{alpha}
\end{equation}
where $r$ is the grains radius. As the crystallites in sample~1 present an average radius of $\sim5\;\mu$m, much larger than $\lambda_{ab}(0)$, $\alpha$ is almost temperature independent up to very close to $T_c$ and was approximated by its low-temperature value ($\sim0.83$). In the case of sample~2, the average grains radius is $\sim 1\;\mu$m, which leads to $\alpha\approx0.35$ when $T\to 0$~K, and decreases monotonously to 0 at $T_c$. In this case, for the sake of simplicity we have used a temperature-independent average value ($\alpha=0.15$). 

ii) In the \textit{London region} of the \textit{mixed state}, i.e., for $\ln\kappa/2\kappa^2<h/(1-t)\stackrel{<}{_\sim}0.3$, it may be approximated\cite{nota}
\begin{equation}
m=-c_1\frac{1-t}{4\kappa^2}\ln\left(\eta\frac{1-t}{h}\right), 
\label{mLondon}
\end{equation}
where $c_1\approx0.77$ and $\eta\approx1.44$ (see e.g., Ref.~\onlinecite{Hao}).

iii) In the \textit{Abrikosov region} of the mixed state, i.e., for $0.3\stackrel{<}{_\sim}h/(1-t)<1$, it is approximated\cite{nota}
\begin{equation}
m=\frac{h-1+t}{\beta_A(2\kappa^2-1)},
\label{mAbrikosov}
\end{equation}
where $\beta_A\approx1.16$ for a triangular vortex lattice. 

iv) In the \textit{normal state} $h/(1-t)>1$, due to the high anisotropy and the relatively high critical temperature of La$_{2-x}$Sr$_x$CuO$_4$ system, thermal fluctuation effects are observable mainly around the $H_{c2}(T)$ line.\cite{tres,Suenaga} However, in the presence of a $T_c$ distribution, these effects may be completely masked by the much larger contribution coming from domains in the fully superconducting state. In particular, in the case of a Gaussian $T_c$ distribution, it was shown that $T_c$ inhomogeneity effects are dominant up to $\sim\overline T_c\pm2\Delta T_c$ (see e.g., Ref.~\onlinecite{trece1}). Then, to study the magnetic behavior in a temperature region $\Delta T_c$ wide about $\overline T_c$, we have discarded any contribution to $m(t,h)$ coming from superconducting fluctuations.  

Finally, as in some cases we are dealing with wide $T_c$ distributions, we have taken into account the $H_{c2}(0)$ dependence on $T_c$. For that, we have evaluated $H_{c2}(0)=\phi_0/2\pi\mu_0\xi_{ab}^2(0)$ with $\xi_{ab}(0)[{\rm nm}]\approx-1.5+130/T_c$[K], as follows from the analysis of the fluctuation diamagnetism above $T_c$ in Ref.~\onlinecite{catorce} (see the inset of Fig.~8 in that reference). However, we have kept constant the other superconducting parameter involved, $\kappa\equiv\lambda_{ab}(0)/\xi_{ab}(0)\approx60$ (see Ref.~\onlinecite{trece1} and references therein). This last approximation is based on the fact that $\lambda_{ab}(0)$ is proportional to $p^{-1/2}$, where $p$ is the hole concentration, and decreases with $T_c$ in the underdoped region roughly in the same way as $\xi_{ab}(0)$ does. 

The comparison with the measurements of the isothermal magnetization curves around $\overline T_c$ calculated on the grounds of the inhomogeneity approach developed above [Eqs.~(4-9)], is summarized in Figs.~\ref{sample1} and \ref{sample2}. For sample 1, Eq.~(\ref{Integralangular}) was evaluated numerically by assuming a Gaussian $T_c$ distribution 
\begin{equation}
\delta(T_c)=c_2\exp\left[-\left(\frac{T-\overline T_c}{c_3\Delta T_c}\right)^2\right],
\label{Distribucion}
\end{equation}
where $c_2$ is a normalization factor and $c_3=1/2\sqrt{\ln{2}}$ according to the above $\Delta T_c$ definition (FWHM).  For sample 1,  it was used $\overline T_c\approx 28.5$ K,  and $\Delta T_c\approx3$ K. This last value is well within the measured $\Delta T_c\approx1.2$ K and $T_c^{\rm max}-\overline T_c\approx3.2$~K (see Table I). As may be seen in Fig.~\ref{sample1}, the resulting isothermal curves are in good agreement with the experimental data, including the $H_{up}$ decrease on increasing the temperature. This last behavior, similar to that of $H_{c1}$, is a direct consequence of the symmetric Gaussian $T_c$-distribution given by Eq.~(\ref{Distribucion}). 

The comparison of the inhomogeneities model with the data for sample 2 is presented in Fig.~\ref{sample2}. The lines in Fig.~\ref{sample2}(b) were again numerically evaluated from Eq.~(\ref{Integralangular}), but this time by using the asymmetric $T_c$ distribution shown in the inset. This last was in turn obtained through the superposition of two Gaussian distributions: one 7 K wide (FWHM) centered in the average $T_c$ (25.5 K), and other 3 K wide centered in a higher $T_c$ (36 K). The volume fraction associated to the higher $T_c$ part represents only 0.8\% of the total volumen fraction. The use of such a distribution is justified by the asymmetric low-field diamagnetic transition found for sample 2 (Fig.~\ref{tc}). As may be appreciated in Fig.~\ref{sample2}, the final result is in good agreement with the experimental data for sample 2, including the $H_{up}$ increase with the temperature. To further check that the behavior of $H_{up}$ observed in sample 2  is associated with  the asymmetry of its  $T_c$ distribution, we have repeated the same calculations of  the magnetization isotherms,  but now removing the higher-$T_c$ bump of the distribution. The results are represented in Fig.~\ref{sample2}(c), where it may be clearly seen that now $H_{up}$ decreases with the temperature.

\section{INSIDE THE FULL INHOMOGENEOUS REGION: GIANT AND NON-LINEAR PRECURSOR DIAMAGNETISM UNDER VERY LOW APPLIED MAGNETIC FIELDS}

An aspect particularly interesting of our magnetization measurements is the diamagnetic behavior observed around $\overline T_c$ and under low field amplitudes, $H\stackrel{<}{_\sim}H_{up}$. The data in Figs.~\ref{sample1}(a) and \ref{sample2}(a) already indicate that the excess-magnetization isotherms measured in the two samples studied present in this region a non-linear dependence with the applied magnetic field.  This behavior, observed in different cuprate superconductors and extensively studied by various groups in different HTSC,\cite{nueve,trece1,trece2,trece3,catorce,quince,dieciseis} was explained in the case of the LSCO-01 samples in terms of $T_c$-inhomogeneities in Sect.~III.B. However, due to the relevance given to these low-field anomalies by different authors,\cite{nueve,quince,dieciseis,diecisiete,dieciocho,diecinueve} they will be further analyzed here by combining the GGL approach with our simple inhomogeneity scenario, and also through new measurements of the reduced-temperature dependence of this anomaly. Another new aspect here will be a throughout comparison between the $\Delta M(T)_H$   and the  $\Delta M(H)_T$  results, which will provide a further check of both the consistency of our experimental results and of the origin of the observed anomalies. 

%
%
% fig7
%
%
\begin{figure}[b]
\includegraphics[scale=0.31]{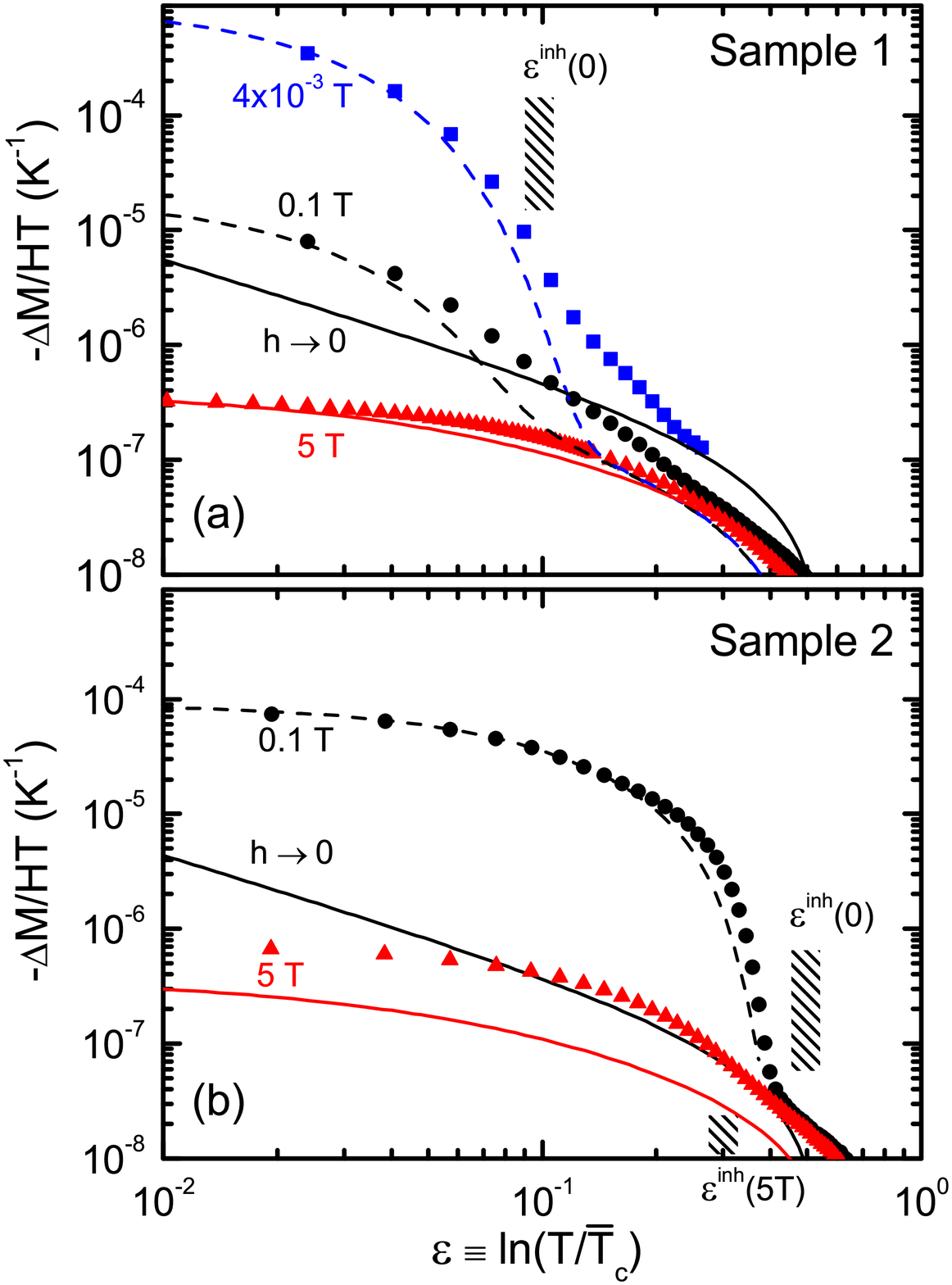}
\caption{(Color online) (a) and (b) Some examples of the excess magnetization (normalized by $HT$) as a function of the reduced temperature. The solid lines are the GGL predictions, obtained from Eq.~(\ref{Prange}) as described in the main text. The dashed areas were calculated from Eq.~(\ref{Criterio}) and correspond to the field-dependent reduced-temperatures below which the measurements are expected to be deeply affected by $T_c$-inhomogeneities. The dashed lines were estimated from the $T_c$-inhomogeneity approach developed in Sect.~III.B. The agreement with the low-field measurements is quite good. As one may also appreciate, in the low field data some inhomogeneity effects manifest  even well above $\varepsilon^{\rm inh}(0)$, up to reduced temperatures of the order of $2\ln(T_c^{\rm max}/\overline T_c)$.  Note also that  in  sample 1, affected only by intrinsic-like inhomogeneities associated with the unavoidable random distribution of Sr ions, a 5 T magnetic field is enough to shift $\overline T_c(H)$ to lower temperatures by more than $2\ln(T_c^{\rm max}/\overline T_c)$.  This  explains why the inhomogeneity effects become non-observable in that case, the agreement with the GGL predictions being then excellent. However, for sample 2 the effect of $T_c$ inhomogeneities extends to much higher reduced temperatures, up to $\varepsilon^{\rm inh}(0)\approx0.5$, and an applied magnetic field  of 5 T  only shifts  $\varepsilon^{\rm inh}(H)$ down to $\sim0.4$. This explains in turn the strong disagreements with the theory of homogeneous samples observed for all field amplitudes in (b).}
\label{DMHTexcess}
\end{figure}

A detail, in linear scale, of the low-field dependence of $\Delta M(H)_T$  measured slightly above $\overline T_c$ is presented in Figs.~\ref{Lineal}(a) and (b) for the two samples studied here.  These experimental isotherms show a pronounced non-linear behavior close to the one of a power-law when $H\to0$,
\begin{equation}
\Delta M(H)_T = AH^z,   
\label{Power}
\end{equation}
where $A$ is a constant and $z$ a \textit{critical} exponent (both of them temperature dependent). The dashed lines in these figures are fits to Eq.~(\ref{Power}), and lead to $z$ values between 0.55 and 0.75. 
This effect was claimed to be an evidence of the unconventional nature of the superconducting transition of these materials.\cite{nueve,quince,dieciseis,diecisiete,dieciocho,diecinueve} Let us note already here that such a behavior, first stressed by Lascialfari and coworkers,\cite{nueve}  must not be confused with a similar   dependence observed above $T_c$  but under much higher reduced magnetic field amplitudes,  in the so-called finite field (or Prange) region (when  $h\approx\varepsilon$, see e.g., Refs.~\onlinecite{dos,tres}). This last behavior  may be easily explained on the grounds of the GGL approach in presence of a total-energy cutoff.\cite{tres,seis} 

As suggested already by the results of Sect.~III.B, the non-linear magnetization behavior at low field amplitudes shown in Figs.~\ref{Lineal}(a) and (c) may be easily accounted for on the grounds of the simple inhomogeneity model developed in  that Section: by using again in  Eq.~(\ref{Integralangular}) the $T_c$ distributions indicated in the insets of Figs.~\ref{sample1}(b) and \ref{sample2}(b), one obtains the solid curves represented in Figs.~\ref{Lineal}(b) and (d). In spite of the approximations introduced in this model, it is remarkable that it reproduces both the observed $\Delta M(H)_T$ amplitude and non-linear behavior. 

To analyze now these low field magnetization anomalies in terms of the GGL approach, one may start by comparing at a quantitative level the excess-magnetization shown in Figs.~\ref{Lineal}(a) and (c) with the GGL predictions. For single-layered superconductors, the excess magnetization in the GGL approach under a total-energy cutoff is given by:\cite{tres}
\begin{eqnarray}
\Delta M=-f\frac{k_BT}{\phi_0s}\left[-\frac{\varepsilon^c}{2h}\psi\left(\frac{h+\varepsilon^c}{2h}\right)-\ln\Gamma\left(\frac{h+\varepsilon}{2h}\right)\right.\nonumber \\
+\left.\ln\Gamma\left(\frac{h+\varepsilon^c}{2h}\right)+\frac{\varepsilon}{2h}\psi\left(\frac{h+\varepsilon}{2h}\right)+\frac{\varepsilon^c-\varepsilon}{2h}\right],~~
\label{Prange}
\end{eqnarray}
where $\Gamma$ and $\psi$ are the gamma and digamma functions, $k_B$ the Boltzmann constant, $\phi_0$ the flux quantum, $s$ the periodicity length of the superconducting layers, $\varepsilon^c$ the total-energy cutoff constant, and $f$ the so-called \textit{effective superconducting fraction},\cite{dos} that may be approximated by the Meissner fraction.\cite{EPLmosqueira}
In the low field regime ($h\ll\varepsilon$) Eq.~(\ref{Prange}) reduces to
\begin{equation}
\Delta M=-f\frac{k_BT}{6\phi_0s}h\left(\frac{1}{\varepsilon}-\frac{1}{\varepsilon^c}\right),  
\label{Schmidtcutoff}
\end{equation}
which corresponds to the Schmidt and Schmid limit\cite{treinta} under a total-energy cutoff. If, in addition, $\varepsilon\ll \varepsilon^c$:
\begin{equation}
\Delta M=-f\frac{k_BT}{6\phi_0s}\frac{h}{\varepsilon}~,  
\label{Schmidt}
\end{equation}
which is the well-known Schmidt-like expression for a single-layered superconductor in the 2D limit.\cite{treinta} 

In the low field regime, the GGL approach predicts, as it is well known, a field-independent excess-magnetic susceptibility, $\Delta\chi\equiv\Delta M/H$.  In addition, by using in Eq.~(\ref{Schmidt}) $f=1$, $s= 6.6$ \r{A}, $T\approx 30$~K and  $\mu_0H_{c2}(0)=36$~T,\cite{treintayuno} one finds $\Delta\chi\approx -2\times10^{-6}\varepsilon^{-1}$,  which takes values orders of magnitude smaller than those that one may directly infer from the experimental data given in Figs.~\ref{Lineal}(a) and (c). For instance, for the isotherms at 30~K in sample 1 and 27~K in sample 2, which roughly correspond to a quite similar reduced temperature of the order of  $\varepsilon \sim 0.05$, and for $\mu_0H=10^{-3}$~T,  the data of Figs.~\ref{Lineal}(a) and (c) lead to $\Delta \chi$  of the order of $-1.3\times10^{-2}$  and $-2.7\times10^{-2}$ for, respectively, samples 1 and 2. These values are to be compared with $\Delta\chi\approx -3.5\times 10^{-5}$, a value directly obtained by using $\varepsilon=0.05$ in Eq.~(\ref{Schmidt}).

To analyze in terms of the GGL approach the temperature dependence of the excess magnetization, one must first extract the experimental $\Delta M(\varepsilon)_H$ by determining the background magnetization and then using Eq.~(\ref{Back}). 
The background magnetization was obtained by fitting a Curie-like function ($M_B = A+BT+C/T$, being $A$, $B$ and $C$ fitting constants) in temperature ranges $\sim 50$ K wide above at least 85 K. Some examples of the magnetization background for each sample may be seen in the insets of Fig.~\ref{isotherms} (corresponding to $\mu_0H = 0.1$~T).
Other aspects of the background extraction are similar to those detailed in Refs.~\onlinecite{trece1} and \onlinecite{catorce}. 

The $\Delta M(\varepsilon)_h/HT$ data of Figs.~\ref{DMHTexcess}(a) and (b) already show, by just comparing to each other the measurements in the two samples, that some of these experimental curves are in fact extrinsic: For instance, at $\varepsilon\sim0.1$  the data measured under a field amplitude of 0.1 T  are almost two orders of magnitude larger in sample~2 than in sample~1. The extrinsic character of most of these $\Delta M(\varepsilon)_h/HT$ curves is further confirmed when they are compared with the GGL predictions: the solid lines in Figs.~\ref{DMHTexcess}(a) and (b) were obtained by using in Eq.~(\ref{Prange}) the parameter values of Table~I, $\varepsilon^c= 0.55$ (see Ref.~\onlinecite{treintaydos}), and $f = 0.95$ and $0.75$ for, respectively, samples~1 and 2.\cite{notaf} These GGL curves strongly differ from the measurements, mainly in the low field and in the low  reduced temperature region, even in the case of sample~1 (affected only by intrinsic-like inhomogeneities associated with the random distribution of the Sr ions).  However, the differences are considerably quenched, even completely in the case of sample~1, for the measurements performed under 5 T (see below). 

The strong disagreement between the GGL approach and the measurements may be easily explained at a qualitative level by taking into account the presence of $T_c$-inhomogeneities. Note first that the experimental $\Delta M(\varepsilon)_h/HT$ curves show a rapid and strong amplitude  increase, and the differences with the GGL predictions are more important at reduced temperatures where the measurements are expected to be deeply affected by $T_c$-inhomogeneities:\cite{trece1}  In the case of $T_c$-inhomogeneities with a Gaussian $T_c$ distribution,  the upper limit of this $\varepsilon$-region, that we will call \textit{full inhomogeneous region}, may be roughly estimated as,\cite{catorce}
\begin{equation}
\varepsilon^{\rm inh}(H)\approx2\frac{\Delta T_c}{\overline T_c}-\frac{H}{H_{c2}(0)}.
\label{Criterio}
\end{equation}            
However, in presence of inhomogeneities that do not follow a Gaussian distribution, the corresponding effects may manifest up to  $T_c^{\rm max}$ (see Section II).

The qualitative results summarized above apply remarkably well to the data shown in Figs.~\ref{DMHTexcess}(a) and (b). The dashed bars here correspond to $\varepsilon^{\rm inh}(H)$ calculated by using the parameters in Table I.  As expected, in the full inhomogeneous region the $\Delta M(\varepsilon)_h/HT$ curves show, for both samples, huge differences with the GGL predictions. But in addition, one may appreciate,  mainly in the case of sample 1, that the data measured under low fields are somewhat affected by inhomogeneities even well above $\varepsilon^{\rm inh}(H)$, up to reduced temperatures of the order of  $2\ln(T_c^{\rm max}/\overline T_c)\approx0.2$. These results are particularly interesting because  they show that the presence of unavoidable, \textit{intrinsic like},  $T_c$-inhomogeneities, even when they have relatively small amplitudes, could have  huge effects on the measurements, even beyond $\varepsilon^{\rm inh}$. The strong sensitivity of the precursor diamagnetism to the presence of $T_c$- inhomogeneities, together with the fact that their  doping make most of the  HTSC intrinsically inhomogeneous, would explain why an anomalous diamagnetism was observed above $T_c$ in a wide number of HTSC families and doping levels.\cite{nueve,diez,once,doce,trece1,trece2,trece3,catorce,quince,dieciseis,diecisiete} Note also that, although these inhomogeneity  effects were already predicted up to $\varepsilon^{\rm inh}$ in Ref.~\onlinecite{trece1} (see the Fig.~6 therein), our present results provide a first experimental evidence of their relevance beyond the full inhomogeneous region. These conclusions  must be probably extended to the behavior of any magnitude around a superconducting transition.

%
%
% fig11
%
%

To further analyze in the full inhomogeneous region how the $T_c$-inhomogeneities affect the temperature dependence of the magnetization curves measured under low field amplitudes, in Figs.~\ref{DMHTexcess}(a) and (b) we have also compared these data with the inhomogeneity approach summarized in Sect.~III.B. The dashed curves in these figures were obtained from Eq.~(\ref{Integralangular}) by including Eq.~(\ref{Schmidtcutoff}) as a contribution to the magnetization in the normal state. 
In doing that, we used for sample 1 a Gaussian $T_c$ distribution with $\overline T_c(0)=28.8$~K and $\Delta T_c=2.5$~K (close to the one used in Sec.~III.B), while for sample 2 we used the double Gaussian distribution used in that Section. In addition, the finite field  effects are crudely implemented in Eq.~(\ref{Schmidtcutoff}) by just taking into account the $T_c$ dependence on $H$. The divergence at $\overline T_c(H)$ is avoided by cutting off the magnetization to its value at the Levanyuk-Ginzburg reduced-temperature, $|\varepsilon_{LG}|\approx 3\times 10^{-2}$. It can be seen how the model developed, despite the crudeness of the approximations used, roughly fits the experimental data using a $\Delta T_c$ just slightly larger than the one determined from low magnetic field susceptibility measurements (see Fig.~\ref{tc}).

Another remarkable result that may be observed in Fig.~\ref{DMHTexcess}(a) is that in sample 1, the one just affected by intrinsic-like  inhomogeneneities (that will be unavoidable even in an ideal LSCO crystal), an applied field of 5 T is enough to shift $\overline T_c(H)$ to low temperatures, making $H/H_{c2}(0)$ even larger than $2\ln(T_c^{\rm max}/\overline T_c)$, [i.e., $\varepsilon^{\rm inh}(H)$ vanishes]. This explains why the inhomogeneity effects become almost non-observable in this case, the agreement with the GGL predictions being excellent in all the studied $\varepsilon$-region. However, for sample 2 its (extrinsic) full inhomogeneous region extends to much higher reduced-temperatures, up to $\varepsilon^{\rm inh}(0)\approx0.5$, and  a field  of 5 T  only shifts $\varepsilon^{\rm inh}(H)$ to around 0.30. This explains the strong disagreements with the theory observed for all field amplitudes in Fig.~\ref{DMHTexcess}(b).

Note finally that the comparison between the $\Delta M(\varepsilon)_H$ curves in Fig.~\ref{DMHTexcess} with the $\Delta M(H)_T$ curves in Figs.~\ref{sample1} and \ref{sample2} provides an important test of consistency for both the magnetization measurements and the procedures followed to estimate the corresponding excess diamagnetism. For instance, the excess magnetization which may be obtained in Fig.~\ref{DMHTexcess}(a) from the data points at $\varepsilon\approx 0.04$ of the $\Delta M(\varepsilon)_H$ curve measured under a magnetic field of $4\times 10^{-3}$ T, is in reasonable agreement with the one that may estimated from the $\Delta M(H)_T$ curve  of Fig.~\ref{sample1}  measured under the same $T$ ($\sim$30 K) and $H$ values.

\section{SUMMARY AND CONCLUSIONS}

The central aim of this work was to contribute to establish the intrinsic precursor diamagnetism in high-$T_c$ superconductors and the origin of the anomalies recently observed in the magnetization above the superconducting transition in underdoped cuprates, in particular in LSCO compounds. These new anomalies include a diamagnetism with a huge and field-dependent amplitude under very low reduced-magnetic fields and, simultaneously, close to the average $T_c$. They induce also an increase when the temperature increases of the so-called upturn magnetic field $H_{up}$.\cite{quince,dieciseis} Moreover, as first reported by Lascialfari and coworkers\cite{quince}, below $H_{up}$ the isothermal magnetization curves show an anomalous power-law dependence on $H$.\cite{dieciseis}
In spite of previous works warning that these type of anomalies could be due to the presence of $T_c$ inhomogeneities associated with chemical disorder,\cite{trece1,trece2,catorce} they are being considered by different authors as intrinsic to underdoped cuprates  and then, without any serious confrontation with other possible causes,  as an evidence for phase incoherent superconductivity in the pseudogap regime.\cite{diecisiete,dieciocho,diecinueve} In fact, these warnings have been taken into account only by Lascialfari and coworkers,\cite{quince} but the inhomogeneity origin of these new magnetization anomalies was discarded because symmetric $T_c$ distributions, the only ones taken into account by these authors, cannot explain the observed temperature dependence of $H_{up}$.

Here, we have first presented detailed experimental results on the magnetization measured above the average transition temperature in two underdoped granular samples of the same nominal composition, La$_{1.9}$Sr$_{0.1}$CuO$_4$ (LSCO-0.1), but with different structural and chemical disorder due to differences in their synthesis. In one of the samples the chemical disorder is just the intrinsic-like one, associated with the unavoidable random distribution of the Sr ions (which will be then present even in an ideal La$_{1.9}$Sr$_{0.1}$CuO$_4$ crystal), and the corresponding $T_c$-inhomogeneities may be approximated as symmetric. In contrast, the most disordered  sample presents a pronounced  asymmetric  $T_c$-distribution. The anomalies observed in the magnetization measurements around the superconducting transition, in particular in the isotherm magnetization curves under low magnetic field amplitudes, are similar to those claimed as intrinsic by other authors. However, the comparison of the as-measured data in both samples already shows directly that most of the observed magnetization anomalies are in fact extrinsic. When analyzed at a more quantitative level, these data first show that these anomalies may be explained by the presence of structural and $T_c$-inhomogeneities with long characteristic lengths, much larger than the in-plane superconducting coherence length amplitude. This conclusion includes the increase of $H_{up}$  as the temperature increases, that may be still attributed to $T_c$-inhomogeneities if they have a markedly asymmetric $T_c$-distribution, extended well above the average transition temperature. In fact, this is mainly the case in the most inhomogeneous samples, where such a temperature dependence of  $H_{up}$ is observed to be more important.

It was also shown experimentally that the inhomogeneity and the fluctuation effects in sample 1, whose chemical disorder is close to the intrinsic one (associated with the unavoidable random distribution of the Sr ions),\cite{catorce} may be disentangled by just applying a magnetic field strong enough as to shift $\overline T_c(H)$ to low temperatures, by an amount  of the order of or larger than the superconducting transition width associated with inhomogeneities.
Under such a field amplitude, for sample~1 of the order of 3~T,  the inhomogeneity effects on the magnetization were quenched, and the remaining diamagnetic effects were explained  in terms of the Gaussian-Ginzburg-Landau approach for homogeneous layered superconductors. This agreement also confirms our previous conclusions about the absence in the bulk of intrinsic inhomogeneities with short characteristic lengths, as those observed by using surface probes.\cite{veinticuatro}

Complementarily, the  agreement with the GGL approach was extended to high reduced-temperatures, for $\varepsilon \stackrel {>}{_\sim} 0.1$, by using a total-energy cutoff   of the order of $\varepsilon^c \sim 0.55$, this last corresponding to the limit imposed by the uncertainty principle to the shrinkage of the superconducting wavefunction when the temperature increases.\cite{treintaydos} This result further confirms our earlier conclusions, obtained through measurements of both the diamagnetism above $T_c$,\cite{tres,trece1,trece2,trece3,catorce} and of the in-plane paraconductivity,\cite{treintaytres} that the onset temperature, $T^c$, for the superconducting fluctuations in underdoped cuprates is not affected by the opening of a pseudogap in their normal state.  Indirectly, these last results also support recent proposals that the large Nernst signal observed in the normal state in LSCO is not associated with superconducting fluctuations.\cite{veintiuno,veintidos} Nevertheless, the relevance that is being given at present to the precise location of  $T^c$ in underdoped cuprates, and to the seeming disagreements between the values inferred from transport or magnetic measurements,\cite{once,quince,dieciseis,diecisiete,dieciocho,diecinueve,veinte,treintaycuatro} make particularly desirable to extend our present magnetization measurements to the high reduced-temperature regime of cuprates with different dopings. For this task, our present results suggest the way to separate the intrinsic fluctuation effects from those due to a, symmetric or not, $T_c$ distribution. 

Another aspect that will need further examination on the grounds of our present results concerns the magnetization measurements above $T_c$ in iron pnictides: whereas detailed results in a high-quality Ba$_{1-x}$K$_x$Fe$_2$As$_2$ single crystal have been explained in terms of conventional GGL approaches,\cite{PRBpnictido} recent measurements under very low field amplitudes in a polycrystalline sample (SmFeAsO$_{0.8}$F$_{0.2}$) present anomalies similar to the ones described here for LSCO samples.\cite{arxivprando}

\section*{Acknowledgments}

We thank Noelia Cot\'on for the preparation of sample 2, Ram\'on I. Rey for the SEM images, and Manuel V. Ramallo for his careful reading of the manuscript and for his useful comments and suggestions. This work was supported by the Spanish MICINN and ERDF \mbox{(FIS2010-19807)}, and the Xunta de Galicia (2010/XA043 and 10TMT206012PR).


\begin{references}
\bibitem{nota}For the sake of simplicity we did't consider possible finite-size effects in the mixed-state magnetization. However, the main contribution to the effective magnetization comes from domains in the Meissner region, where these effects were taken into account.
\bibitem{notaf}In the present case in which we are using randomly oriented polycristalline samples, the Meissner fraction was estimated as the ratio between the field-cooled and zero-field-cooled magnetic susceptibilities at low temperatures.
\end{references}
\end{document}